\documentclass{aa}   
\usepackage{graphics}

\begin{document}

\title{The effect of an attenuated continuum on the coronal line
spectrum of NGC~1068 and the Circinus galaxy.} 
\author{L.S.Nazarova\inst{1,2},
P.T.O'Brien\inst{1},
M.J.Ward\inst{1}
}

\offprints{P.T. O'Brien}

\institute{Department of Physics \& Astronomy, University of Leicester,
University Road, Leicester, LE1 7RH, U.K.
\and 
Astronomical Society, Sternberg Astronomical Institute, Universitetskij 
prosp.13, Moscow, 119899, Russia}

\thesaurus{11(11.01.2; 11.17.2; 02.18.7; 11.09.1 NGC~1068, Circinus)}

\date{Received/ Accepted}

\authorrunning{Nazarova et al.}
\titlerunning{The coronal lines in NGC~1068 and the Circinus galaxy}

\maketitle 

\begin{abstract}
We present photoionization models of the optical and IR coronal line
spectrum in NGC~1068 and the Circinus galaxy. The line fluxes have been
calculated using (a) a non-thermal (nuclear) continuum source and (b) the
non-thermal continuum plus a UV bump due to a stellar cluster. We take
into account the effect of attenuation of these continua by gas with
column density 10$^{22}$ cm$^{-2}$ located between the nucleus and the
coronal line region. The calculated coronal line ratios are in a good
agreement with those observed in NGC~1068 for a model in which about 40\%
of the line emission comes from gas illuminated by unattenuated,
non-thermal continuum, and about 60\% from gas illuminated by attenuated,
non-thermal continuum. The electron density of the coronal line emitting
gas in NGC~1068 is found to be 10$^{4}$ cm$^{-3}$. In the Circinus galaxy
the coronal line emission comes from gas with electron density 10$^{3}$
cm$^{-3}$ illuminated entirely by attenuated, non-thermal continuum. The
derived ionization parameters for both coronal line regions are very
similar, but the different densities imply a higher ionizing photon flux
in NGC~1068, consistent with the higher observed excitation state of the
line emitting gas in that galaxy. A possible geometry of the coronal line
region of both galaxies is discussed, in which the distribution of the
attenuating gas may be strongly affected by the relative strength of the
nuclear radio-jet. The stronger radio-jet in NGC~1068 may have cleared a
channel through the NLR allowing some unattenuated nuclear continuum to
illuminate part of the coronal line region.

\end {abstract}

\section{Introduction}

Coronal lines correspond to high-excitation, forbidden transitions of
ionized species with ionization potentials, E $\geq$ 100 eV. The study of
the optical coronal lines --- the so-called iron coronal lines
[\ion{Fe}{vii}]$\lambda$6087, [\ion{Fe}{x}]$\lambda$6375,
[\ion{Fe}{xi}]$\lambda$7892 and [\ion{Fe}{xiv}]$\lambda$5303 --- has a
long history beginning with the original work of Seyfert (1943). Recently
interest in the subject has revived following the detection of near--IR
coronal lines (Oliva \& Moorwood 1990; Oliva et al. 1994). Infrared
observations of NGC~1068 and the Circinus galaxy show the presence of
strong coronal line emission, such as
[\ion{Si}{vi}]$\lambda$1.963($\mu$m),
[\ion{Si}{vii}]$\lambda$2.483($\mu$m), [\ion{S}{viii}]$\lambda$9912,
[\ion{S}{ix}]$\lambda$1.25235($\mu$m),
[\ion{Si}{ix}]$\lambda$3.934($\mu$m) and
[\ion{Si}{x}]$\lambda$1.4305($\mu$m), (Oliva et al. 1994; Marconi et al.
1996; Moorwood et al. 1996; Thompson 1996; Oliva 1997). The energy needed
to produce these ions is in the range 167--351 eV. The optical iron
coronal lines have ionization potentials of 100--361 eV. Thus, modelling
of these coronal lines can provide important information concerning the
shape of the EUV ionizing continuum in the energy range 100--360 eV.

Investigation of the origin of the iron coronal lines show that they are
not fitted well by models which assume gas either purely collisionally
ionized by fast shocks or photoionized by hot stars (Viegas-Aldrovandi \&
Contini 1989; Oliva et al. 1994; Marconi et al. 1996). For example, the
observed [\ion{Si}{vi}]/[\ion{Fe}{vii}] ratio in NGC~1068 is better
predicted by photoionization models (Marconi et al. 1996). Although
Marconi et al. did not exclude some contribution to the coronal line
emission in NGC~1068 from shocks or X-rays emitted by a shock front, they
suggested that the coronal line region is predominantly photoionized.
Recent long-slit, HST observations of NGC~1068 obtained by Axon et al.
(1998) (closest spectrum to the proposed nucleus is at $\sim 1.5\arcsec$)
have shown that some of the strong [\ion{Fe}{vii}]$\lambda$3769 emission
coincides with the radio-jet axis. In the Circinus galaxy shock heating
is effectively excluded because the coronal line widths are less than 100
km s$^{-1}$.

The most likely source of excitation of the coronal lines is
photoionization by the hard UV continuum from a ``bare'' Seyfert nucleus
(Korista \& Ferland 1989; Oliva et al. 1994; Marconi et al. 1996;
Ferguson et al. 1997). Ferguson et al. found that the coronal line gas is
likely to be dust free and that the ionization parameter, U, and the gas
electron--density, N$_{e}$, and temperature, T$_{e}$, in the coronal line
region are $-2.0 \leq$ Log~U $\leq 0.75$, $10^{2} \leq$ N$_{e} \leq
10^{8.5}$ cm$^{-3}$ and 12 000 $\leq $ T$_{e} \leq$ 150 000 K
respectively.

In addition to the range of physical parameters investigated by Ferguson
et al. (1997) one more --- the shape of the ionizing continuum --- should
be considered. Indeed, it is difficult to model successfully the coronal
line ratios even in the two best studied galaxies, NGC~1068 and Circinus,
by assuming just different physical conditions in the line emitting gas.
For example, the observed [\ion{Fe}{vii}]/[\ion{Fe}{x}] ratio is
6.3$\pm$0.9 in NGC~1068 and 0.47$\pm$0.08 in Circinus. The
[\ion{Si}{ix}]/[\ion{Si}{vi}] ratio is 0.63$\pm$0.10 in NGC~1068 and
2.12$\pm$0.32 in Circinus (Marconi et al. 1996). It is difficult to
explain such striking differences in these ratios based on
photoionization models of a coronal line region with solar abundances
ionized by a continuum similar to that of a typical Seyfert galaxy
(Ferguson et al. 1997).

In this paper we examine the effect of four different continuum shapes on
the coronal line intensities predicted by photoionization models for
NGC~1068 and the Circinus galaxy. The nuclear continuum, henceforth
referred to as the `non-thermal continuum' to distinguish it from any
stellar contribution (discussed below), arises from the unresolved,
nuclear region around the central black hole. This nuclear continuum may
itself be partly thermal emission from an accretion disk. We also
consider the possible effect of a strong `UV bump', which could be an
additional ionization source for low-ionization coronal species, such as
[\ion{Fe}{vii}] and [\ion{Si}{vi}], compared to high--ionization species,
like [\ion{Fe}{x}] and [\ion{Si}{ix}]. The most likely explanation for a
UV bump in NGC~1068 is the presence of a young stellar-cluster near to
the nucleus (Thatte et al. 1998). Nazarova (1995) found that the strong
[\ion{Ne}{v}]$\lambda$3425 emission from the ENLR in NGC~1068 could be
explained if the shape of the ionizing continuum is the sum of both
non-thermal and thermal components The Circinus galaxy is also
characterized by circumnuclear starburst activity. Star formation
activity has been traced by H$\alpha$ (Marconi et al. 1994), and the Br
recombination line emission peaks at radius $\approx 200$ pc (Moorwood \&
Oliva 1994).

As the ionization potentials for the coronal lines are located in the
energy range which can be significantly affected by even a relatively
modest amount of continuum attenuation (see Fig. 2), we also
examine the effect of attenuation of the nuclear and nuclear+stellar
continua by gas located between the nucleus and the coronal line region.
Photoionization models for the ENLR in NGC~1068 show that the ENLR may be
illuminated by an attenuated continuum (Evans \& Dopita, 1986; Bergeron
et al. 1989; Binette et al. 1996; Nazarova et al. 1998).

The paper is organized as follows: the adopted parameters of the models
are discussed in section 2; the modelled surface-luminosity of the
coronal line regions of NGC~1068 and the Circinus galaxy for different
lines and for the four different continuum shapes are presented in section~3
and the discussion and conclusions are given in sections 4 and 5
respectively.


\section{The model parameters}

\subsection{Luminosity of the central source}

The observed, electron-scattered nuclear-flux of NGC~1068 in the spectral
range between 10$^{14.6}$ and 10$^{18.4}$ Hz is 1.5$\times$10$^{-10}$ erg
cm$^{-2}$ s$^{-1}$ (Pier et al. 1994). The corresponding luminosity of
$3.5\times10^{42}$ erg s$^{-1}$, adopting a distance of 14.4~Mpc
($1\arcsec =$ 72 pc; Tully 1988), provides a lower limit to the intrinsic
nuclear luminosity. In our previous paper on NGC~1068 (Nazarova et al.
1998), we found that the observed high-excitation ENLR could be
successfully modelled if the intrinsic nuclear luminosity is actually
$3.6\times10^{44}$ erg s$^{-1}$ between 10$^{14.6}$ and 10$^{18.4}$ Hz.
We adopt the same luminosity, which is approximately half the FIR
luminosity of NGC~1068 found by Telesco and Harper (1980). This intrinsic
luminosity implies a scattering efficiency of 1\% , consistent with that
estimated by Pier et al. (1994).

The bolometric FIR luminosity of the Circinus galaxy is $\approx$ $4
\times 10^{43}$ erg s$^{-1}$ (Moorwood \& Glass 1984; Ghosh et al. 1992),
adopting a distance of 4 Mpc ($1\arcsec =$ 20 pc). The observed X-ray
luminosity of the hidden nucleus of the Circinus galaxy is $\approx 4
\times 10^{41}\,\Omega/2\pi$, where $\Omega$ is the solid angle subtended
at the nucleus by the visible part of the reflecting matter (Matt et al.
1996). Moorwood and Glass (1984) and Forbes and Norris (1998) suggest
that the FIR luminosity in Circinus may be dominated by processes
associated with circumnuclear star formation. In contrast, Moorwood et
al. (1996) suggest the starburst contributes about 10\%. Given the
uncertainties in the continuum shape, dust covering factor and in the
isotropy of the ionizing continuum we derive an ionizing luminosity in an
analogous way to that used for to NGC~1068. We assume that the FIR
luminosity of the Circinus galaxy is a upper limit to the ionizing
luminosity, and that approximately half of it could be due to dust heated
by stars. Thus, we adopt an ionizing luminosity for the Circinus galaxy
of $2 \times 10^{43}$ erg s$^{-1}$. An increase or decrease in this
number by (say) 50\% would make no significant difference to the
conclusions reached in this paper. 

\subsection{Location and kinematics of the coronal line region}

\subsubsection{NGC~1068}

The observed coronal line emission in NGC~1068 peaks $\approx
0.5\arcsec$ NE of the nucleus and extends over $\approx 4\arcsec$
(Marconi et al. 1996). This position is close to that found by Marconi et
al. for other similar ionization--potential lines at widely different
wavelengths, such as [\ion{S}{iii}]$\lambda$9069 and
\ion{He}{ii}$\lambda$4686 and [\ion{Fe}{xi}] and [\ion{Si}{ix}].
They also found that low--excitation lines with a ionization potential E
$\leq 20$ eV have a mean central velocity of $\simeq 1080$ km s$^{-1}$,
which is close to the systematic velocity, medium excitation lines with
$20\leq$ E $\leq 100$ eV have a central velocity $\simeq 930$ km s$^{-1}$
and coronal lines with E $\geq 100$ eV have a central velocity $\simeq
850$ km s$^{-1}$. Although this trend suggests decelerating, outflowing
gas, the [\ion{Fe}{xi}] and [\ion{Si}{ix}] lines in NGC~1068 are
significantly narrow than [\ion{Fe}{vii}]. One possible reason is that
[\ion{Fe}{vii}] may have a similar rotational component of velocity as
[\ion{O}{iii}] at the same distance from the central source. Indeed the
blue--shifted component of [OIII] may be mainly emitted by outflowing
material, as it has a velocity-shift similar to the coronal lines,
whereas the velocity of the narrow [OIII] component is closer to the
systematic velocity and is possibly due to rotating material (Alloin
1983). We assume that the coronal line region in NGC~1068 is located
between $0.5\arcsec$ and $4\arcsec$ (36--288 pc). Adopting the continuum
shapes discussed in Section 2.3, the unattenuated, incidental,
monochromatic continuum flux from the central source at the inner radius
of the coronal line region is $\nu$F$\nu$ = 10$^{2.46}$ erg s$^{-1}$
cm$^{-2}$ at 0.1 Ryd.

\subsubsection{The Circinus galaxy}

The nucleus of the Circinus galaxy shows a one-sided ionization cone
(Marconi et al. 1994) and bipolar, polarized radio lobes (Elmouttie et
al. 1995). The size of the coronal region emitting the highest excitation
lines, estimated from the [\ion{Fe}{x}] image, is $\approx 10$ pc
($\approx 0.5\arcsec$) (Oliva et al. 1994). Other coronal lines are also
highly concentrated towards the nucleus. The visible and IR coronal lines
are blue-shifted by about 35 km s$^{-1}$ in Circinus (Oliva et al.). As
the luminosity of the Circinus galaxy is $\approx 18$ times lower than
NGC~1068 and the adopted inner-radius of the coronal line region in
NGC~1068 is 36~pc, we adopt an inner radius for the coronal line region
in the Circinus galaxy of 8.6~pc. This convenient definition means that
the unattenuated, incidental, monochromatic continuum flux from the
central source at the inner radius of
the coronal line region in the Circinus galaxy is the same as in NGC~1068
(i.e., $\nu$F$\nu$ = 10$^{2.46}$ erg s$^{-1}$ cm$^{-2}$ at 0.1 Ryd, for
the continuum shapes discussed below).

\subsection{The shape of the ionizing continuum}

We model the coronal line intensities of NGC~1068 and the Circinus galaxy
using four different ionizing continua. The first is the `non-thermal',
nuclear continuum. It should be noted that we refer to this continuum as
the non-thermal continuum, although it may in fact have a thermal
contribution, simply in order to distinguish it from the continuum
discussed below which includes a hot, possibly stellar, UV bump
component. The shape of the non-thermal continuum for NGC~1068 was
derived from two sources. The UV and X-ray continua were taken from Pier
et al. (1994), extrapolated to the EUV region. The rest of the continuum
shape was taken as that of the canonical AGN continuum of Mathews \&
Ferland (1987). This continuum is the same as that used in our previous
paper modelling the ENLR in NGC~1068 (Nazarova et al. 1998). The shape of
the central continuum for the Circinus galaxy is less well known. The
X-ray radiation in the 2--10 keV band is heavily obscured (Matt et al.
1996) and a compact radio synchrotron source is observed in the nucleus
(Forbes \& Norris 1998). Because of the uncertainties in the shape of the
ionizing continuum in the Circinus galaxy we adopt the same continuum
shape as for NGC~1068 when modelling the coronal line intensities. The
photoionization model with this incident continuum shape is called Model
NT.

\begin{figure}
\resizebox{\hsize}{!}{\includegraphics{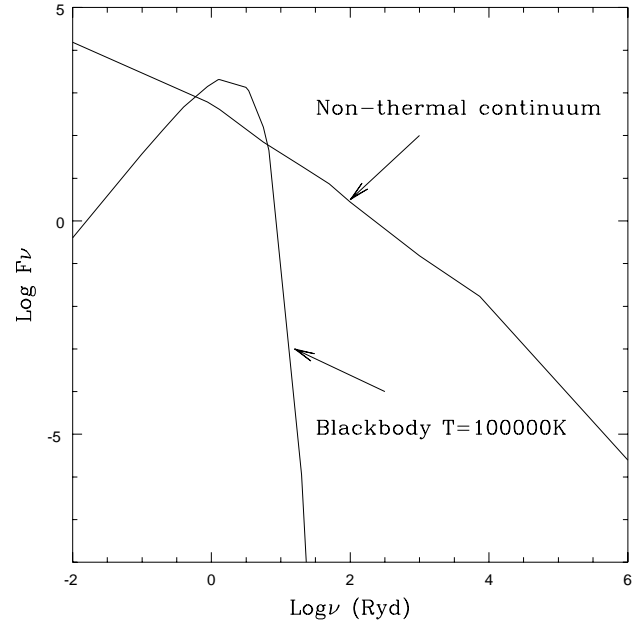}}
\caption{The non-thermal  and stellar (T=100 000K) continua.  
Both continua are plotted in
units of photons Ryd$^{-1}$ cm$^{-2}$ s$^{-1}$ with an arbitrary 
normalization.}
\end{figure}

Based on integral-field spectroscopy, Thatte et al. (1998) suggest that
about 10\% of the bolometric luminosity in the central $1\arcsec$ of
NGC~1068 comes from a moderately--reddened stellar cluster, age 5--13
$\times 10^{8}$ years, intrinsic size $\approx 50$ pc surrounding the
nucleus. However, they note that the stellar cluster could contribute as
much as 50\% of the nuclear bolometric luminosity if the cluster age is
$\leq 10^{7}$ years. An old starburst population is probably present in
the Circinus galaxy, located within the innermost 4 pc (Oliva et al.
1994). Moorwood et al. (1996) modelled the IR coronal lines with a
central ionizing continuum which has a UV bump peaking around 70 eV. The
ASCA data for Circinus are consistent with soft X-ray emission from a
circumnuclear starburst (Matt et al. 1996).

Although we are certain of neither the existence nor origin of a UV
bump in either NGC~1068 or Circinus, we calculated the coronal line
intensities including a `UV bump' component in the incident continuum in
addition to the non-thermal continuum. We assume that the UV bump
component can be represented by a blackbody continuum with temperature
T~$= 100$~000~K, a temperature typical of that of the hottest stars
(the so-called `warmers', Terlevich \& Melnick 1985). The
photoionization model with this combined non-thermal + `stellar' continuum is
called Model NTS. The non-thermal and stellar continuum components are
shown in Fig. 1. The relative contributions to the ionizing flux, as seen
by the coronal line gas, due to the non-thermal and stellar components in
Model NTS are 33\% and 67\% respectively. It is difficult to estimate the
actual ionizing flux due to the star cluster as seen by the coronal line
gas as we cannot be certain of either the distance from the stars to the
gas or the stellar luminosity. Here we have deliberately set the stellar
ionizing flux seen by the coronal line gas at a high value to examine the
effect of such a component. 

To construct the other two continua we included the effect of attenuation
of both the NT and NTS continua. We used the Cloudy photoionization code
(version 90.03; Ferland 1996; Ferland et al. 1998) to calculate the sum
of both the transmitted and diffuse continua for an attenuating gas
column of $10^{22}$ cm$^{-2}$ with N$_{e} = 10^{4}$ cm$^{-3}$, located
between the nucleus and the coronal line region. The photoionization
models with these attenuated continua are called Models NTA and NTSA
respectively. The shapes of the four continua used to model the coronal
line regions in NGC~1068 and the Circinus galaxy are shown in Fig. 2.

Adopting these four continuum shapes, all the models were normalized to
have the same monochromatic flux illuminating the inner-radius of the
coronal line region at 0.1 Ryd as in model NT, namely $\nu$F$\nu$ =
$10^{2.46}$ erg s$^{-1}$ cm$^{-2}$.

\begin{figure}
\resizebox{\hsize}{!}{\includegraphics{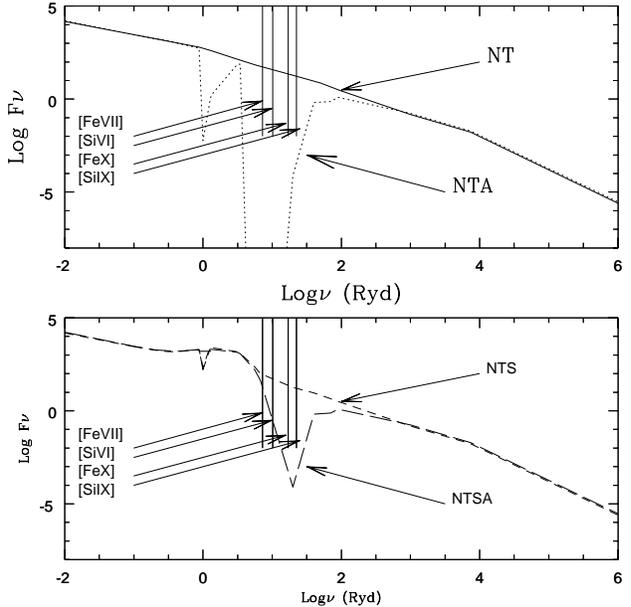}}
\caption{The continua for the four models plotted in
units of photons Ryd$^{-1}$ cm$^{-2}$ s$^{-1}$ with arbitrary
normalization. The photon energies needed to produce several prominent
coronal line species are also indicated.}
\end{figure}

\section{Photoionization models}

We calculated the emission line intensities using the CLOUDY
photoionization code (version 90.03; Ferland 1996; Ferland et al. 1998)
for a slab of gas with solar abundances. Based on the observations, the
inner and outer radii of the coronal line region in NGC~1068 were taken
to be 36 pc ($0.5\arcsec$) and 288 pc ($4\arcsec$) respectively. The size
of the coronal line region in NGC~1068 agrees well with the slit width
($4.4\arcsec$) of the IR spectra (Marconi et al. 1996). For the optical
data, we used the dereddened line fluxes in the range 3400--5100\AA\ and
4800--9300\AA\ given by Marconi et al., which were taken with slit widths
of $1.4\arcsec$ and $1.5\arcsec$ respectively.

The size of the region emitting the high excitation lines in the Circinus
galaxy, estimated from the [\ion{Fe}{xi}] image, is about 10 pc
($\approx 0.5\arcsec$) and the long-slit spectra show that the elongation of
the coronal line region is less than $2\arcsec$ (Oliva et.al., 1994). For
the Circinus galaxy we took the inner-radius of the coronal line region
as 8.6 pc ($0.43\arcsec$) and the outer radius as 68.8 pc ($3.44\arcsec$)
(section 2.2.2).

The coronal line regions in NGC~1068 and the Circinus galaxy are
sufficiently large that we normalize all models to the ionization
parameter U at the inner-edge of the coronal line region. The `local
value' of U will of course vary across the region. The line intensities
used in this paper were computed by integrating across the entire
coronal-line region. The ionization parameter, U, depends on several
factors: the electron density, the flux at a given frequency and the
shape of the ionizing continuum. Although the continuum flux at 0.1 Ryd
illuminating 1~cm$^{2}$ of gas at the inner-radius of the coronal line
region was fixed to be the same in all four models, the different
ionizing continuum shapes result in a different number of ionizing
photons and hence give rise to a different ionization parameter at the
inner-radius.

For ease of visual comparison, the results for the various models
presented in Figs. 3, 4, 5 and 6 are plotted against units of Log~U for
the non-thermal, unattenuated continuum model (Model NT) at the
inner-radius of the coronal line region. The conversion from Log~U for
the other three models to that of Model NT was achieved by adding the
difference in Log~U between the other models and Model NT ($\Delta$Log~U,
which was calculated using the data given in Table~1). For example, for
gas with an electron density of $10^{4}$ cm$^{-3}$ the ionization
parameter for Model NT is Log~U = $-1.06$. However, for Model NTA the
ionization parameter is Log~ U = $-2.4$, and hence $\Delta$Log~U = 1.34.
Although we present all results in units of Log~U for Model NT, when we
derive ``best fit'' values of Log~U we also give the ionization
parameters corresponding to the specific model being considered
calculated from the value of $\Delta$Log~U.

The flux of ionizing photons, N$_{ph}$ (cm$^{-2}$ s$^{-1}$), in each
model together with Log~U at the inner-radius of the coronal line region
for an electron density N$_{e} = 10^{4}$ cm$^{-3}$ are presented in
Table~1. The relation between electron density, N$_{e}$, and Log~U (Model
NT) derived from the model grids for the four different models are shown
in Fig. 3(a). The number of ionizing photons is the same for all electron
densities for the same model, but Log~U varies according to 
changes in the electron density.

\begin{figure}
\resizebox{\hsize}{!}{\includegraphics{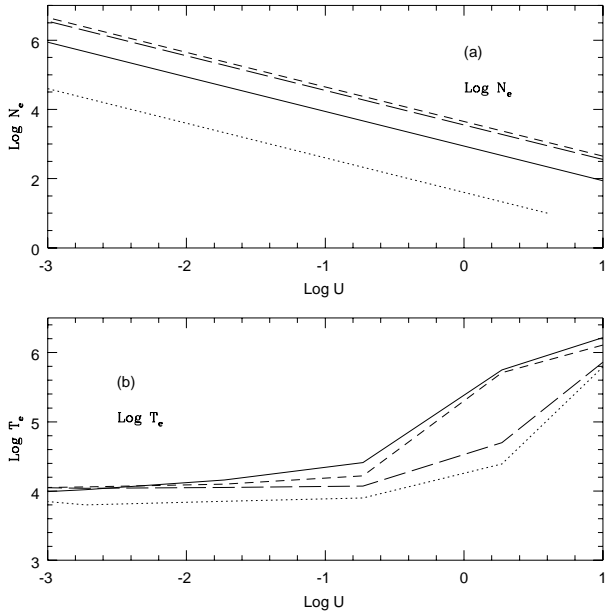}}
\caption{The variation of Log N$_{e}$ and Log T$_{e}$ with Log~U.
The values of Log~U correspond to those for Model NT (see text for
details). The solid line is for Model NT, the small dashed line is for
Model NTS, the dotted line is for Model NTA and the long dashed line is
for Model NTSA.}
\end{figure}

\begin{table*}
\caption{The parameters of the models for  N$_{e} = 10^{4}$ cm$^{-3}$.}
\begin{tabular}{lrrrrr}
\hline
\noalign{\smallskip} 
Models&NT&NTA&NTS&NTSA\\
\hline
       \noalign{\smallskip}
Log N$_{ph}$ & 13.41&12.07&14.12&14.00\\
Log~U&$-$1.06&$-$2.40&$-$0.35&$-$0.45\\
  \noalign{\smallskip}
            \hline
        \end{tabular}
\end{table*}

The derived average electron temperature, T$_{e}$, in the coronal line
region is shown in Fig. 3(b). The variation of T$_{e}$ with Log~U in
Models NT, NTS and NTSA is similar for Log~U $\leq -2$. For Model NTA
when Log~U $\leq -0.3$, T$_{e} \leq 10^{4}$~K. The attenuated Models NTA
and NTSA show smaller increases in temperature with increasing ionization
parameter compared to Models NT and NTS.

\subsection{The coronal lines}

The calculated coronal line intensities are shown in Fig. 4 for each
model with several different electron densities. In Fig. 4 the variation
in the ionization parameter for each specific continuum reflects the
different densities of the emitting gas. Clearly, the shape of the
central continuum strongly affects the lines for all models.
Including attenuation (Models NTA and NTSA) shifts the maximum line
emission to higher ionization parameter compared to that for the unattenuated
continua (Models NT and NTS). This is due to the smaller ionization
parameter for the attenuated continua, as discussed previously (see Table 1).

\begin{figure*}
\resizebox{\hsize}{!}{\includegraphics{7865.f4}}
\caption{The variation in line emission with ionization
parameter for eight coronal lines, plotted in units of erg cm$^{-2}$
s$^{-1}$. The numbers show the photon energies (in eV) needed to produce
the ions of the coronal line parent species. The values of Log~U
correspond to those for Model NT (see text for details). The solid line
shows the emission from 1~cm$^{2}$ of the coronal line gas for Model NT
(unattenuated, non-thermal continuum); the small dashed line shows the
emission for Model NTS (the sum of unattenuated, non-thermal+stellar
continua); the dotted line corresponds to the attenuated, non-thermal
continuum (Model NTA); and the long dashed line corresponds to the attenuated,
non-thermal+stellar continua (Model NTSA).}
\end{figure*}

The calculated high-ionization coronal line ratios for each model are
shown in Fig. 5, together with the observed ratios from Marconi et al.
(1996) for NGC~1068, and from Oliva et al. (1994) for the Circinus galaxy.
As noted, a higher ionization parameter is required to fit the
high-ionization line strengths with the attenuated continua (Models NTA
and NTSA) compared to the unattenuated continua (Models NT and NTS).

\begin{figure*}
\resizebox{\hsize}{!}{\includegraphics{7865.f5}}
\caption{The high-ionization coronal line ratios. The values of Log~U
correspond to those for Model NT (see text for details). The observed
ratios in NGC~1068 are shown by the solid horizontal lines and the
observed ratios in the Circinus galaxy are shown by the dashed horizontal
lines. The photoionization model results are shown by the other lines,
using the same notation as in Fig. 4.}
\end{figure*}


\subsubsection{NGC~1068}

It can been seen from Fig. 5 that model line ratios close to those
observed in NGC~1068 (solid horizontal lines) correspond to different
ionization parameters. For example, the observed
[\ion{Fe}{xi}]/[\ion{Fe}{x}] ratio corresponds to an ionization parameter
Log~U $\approx -1$ for Model NT, whereas the
[\ion{Fe}{vii}]/[\ion{Fe}{x}] ratio corresponds to Log~U $\approx -1.5$.
Similarly, a difference of $\Delta$Log~U $\approx 2$ is found between the
[\ion{S}{ix}]/[\ion{S}{viii}] and [\ion{Fe}{xi}]/[\ion{Fe}{x}] ratios for
Model NT. The [\ion{Si}{ix}]/[\ion{Si}{vi}] and
[\ion{Fe}{vii}]/[\ion{Fe}{x}] ratios correspond to a similar Log~U for
Model NTSA, whereas the [\ion{Fe}{xi}]/[\ion{Fe}{x}] and
[\ion{S}{ix}]/[\ion{S}{viii}]) ratios correspond to very different
ionization parameters.

Because the observed coronal line ratios in NGC~1068 are difficult to
explain with a single ionizing continuum shape, we investigated whether
they could be fitted using a mixture of gas illuminated by unattenuated
and attenuated continua. Line ratios for such a mixture of gas components
are shown in Fig. 6, where the value r = NT/NTA and r = NTS/NTSA denotes
the relative contribution of line emission from the unattenuated and
attenuated models. When r $= 1$ the lines are emitted from the attenuated and
unattenuated models in equal proportion. When r $= 0.7$ the unattenuated and
attenuated models contribute 41\% and 59\%
respectively of the line emission. When r $= 0.5$ the contributions are
33\% and 66\% respectively.

\begin{figure*}
\resizebox{\hsize}{!}{\includegraphics{7865.f6}}
\caption{The coronal line ratios for a combination  of
the attenuated and unattenuated models. The value r = NT/NTA or r =
NTS/NTSA is the relative contribution of line emission from the
unattenuated and attenuated models. The values of Log~U correspond to
those for Model NT (see text for details). When r $= 1$ the lines are
emitted from the attenuated and unattenuated models in equal proportion.
When r $= 0.7$ the relative contributions are 41\% unattenuated and 59\%
attenuated and when r $= 0.5$ the contributions are 33\% unattenuated and
66\% attenuated. The solid line corresponds to the pure non-thermal
continua (NT and NTA) and the long dashed line corresponds to the sum of
non-thermal and UV bump continua (NTS and NTSA).}
\end{figure*}

From Fig. 6, we find the best-fit solution for NGC~1068 for the 3
high--ionization coronal line ratios is with r $= 0.7$ for Models NT
(41\%) and NTA (59\%) and Log~U $= -1$. From Fig. 3(a) this ionization
parameter corresponds to N$_{e} \approx 10^{4}$ cm$^{-3}$. This electron
density is quite close to the N$_{e} \geq 10^{4.5}$ cm$^{-3}$ derived by
Axon et al. (1998) from the [\ion{Ar}{iv}]$\lambda\lambda$4710/4740 line
ratio. The difference in the electron density obtained by Axon et al. and
the electron density estimated by our modelling could be due to the very
different spatial resolutions of the IR/optical (Oliva et al. 1994;
Moorwood et al. 1996) and HST spectra. Axon et al. obtained a spectrum
with a slit projected along the jet and found dense blobs inside of the
jet, whereas the ground-based IR/optical spectra average over the coronal
line region.

The ratio of the observed to modelled (NT+NTA, r $=0.7$) line fluxes as a
function of ionization energy for NGC~1068 are shown in Fig. 7. The model
shows a reasonably good fit for the sulphur, silicon and iron coronal
lines. We note that the ionization parameter, Log~U $= -1$ quoted above
and obtained from the best-fit NTA+NT model, is in units of Log~U for
Model NT. Using the data in Table~1 (Section 3), the ``true'' ionization
parameter for 41\% Model NT + 59\% Model NTA (r $= 0.7$) is Log~U
$\approx -1.4$.

\begin{figure}
\resizebox{\hsize}{!}{\includegraphics{7865.f7}}
\caption{Observed divided by modelled flux ratios in NGC~1068, for the
NT+NTA model with r $= 0.7$.
All ratios are expressed relative to [\ion{Si}{ix}]3.935$\mu$m. The
observed intensities of the lines [\ion{Fe}{vii}]$\lambda$6087,
[\ion{Fe}{x}]$\lambda$6375, [\ion{Fe}{xi}]$\lambda$7892,
[\ion{Fe}{xiv}]$\lambda$5303, [\ion{Si}{vi}]$\lambda$1.963($\mu$m),
[\ion{Si}{vii}]$\lambda$2.483($\mu$m),
[\ion{Si}{ix}]$\lambda$3.9346($\mu$m), [\ion{S}{iii}]$\lambda$9052,
[\ion{S}{viii}]$\lambda$9912 and [\ion{S}{ix}]$\lambda$1.25235($\mu$m)
are taken from Marconi et al. (1996). The observed intensity of the
[\ion{Si}{x}]$\lambda$1.4305($\mu$m) line was taken from Thompson
(1996).}
\end{figure}

We did not find a reasonable solution for NGC~1068 using the
non-thermal+stellar continua (the NTSA+NTS models) for the 3 line ratios
shown in Fig. 6. The [\ion{Fe}{vii}]/[\ion{Fe}{x}] and
[\ion{S}{ix}]/[\ion{S}{viii}] ratios do not strongly depend on the
relative contribution from the attenuated and unattenuated models,
whereas the [\ion{Si}{ix}]/[\ion{Si}{vi}] ratio does depends strongly on
their relative contribution. When r=1 (equal contribution from attenuated
and unattenuated models) the [\ion{Si}{ix}]/[\ion{Si}{vi}] ratio is
shifted to lower ionization parameters for both the NTA+NT and NTSA+NTS
continua. This is due to the decrease in the number of ionizing photons
resulting from the higher relative contribution from regions which are
illuminated by the attenuated continuum. The
[\ion{Si}{ix}]/[\ion{Si}{vi}] ratio thus serves as an indicator of the
relative contribution of gas illuminated by the attenuated and
unattenuated continua.

\subsubsection{The Circinus galaxy}

\begin{figure}
\resizebox{\hsize}{!}{\includegraphics{7865.f8}}
\caption{Observed divided by modelled (Model NTA) flux ratios in the
Circinus galaxy. All ratios are expressed relative to
[\ion{Si}{ix}]3.935$\mu$m. The intensities of the lines
[\ion{Fe}{vii}]$\lambda$6087, [\ion{Fe}{x}]$\lambda$6375,
[\ion{Fe}{xi}]$\lambda$7892, [\ion{Fe}{xiv}]$\lambda$5303,
[\ion{Si}{vi}]$\lambda$1.963($\mu$m),
[\ion{Si}{vii}]$\lambda$2.483($\mu$m),
[\ion{Si}{ix}]$\lambda$3.9346($\mu$m),
[\ion{S}{iv}]$\lambda$10.54($\mu$m) [\ion{S}{ix}]$\lambda$1.25235($\mu$m)
are taken from Oliva et al. (1994). The intensity of the lines
[\ion{S}{iv}]$\lambda$10.54($\mu$m), [\ion{Mg}{v}]$\lambda$7.66($\mu$m),
[\ion{Mg}{vii}]$\lambda$5.51($\mu$m) and
[\ion{Mg}{viii}]$\lambda$3.03($\mu$m) are taken from Moorwood et al.
(1996).}
\end{figure}

Unlike NGC~1068, we were able to find a reasonable fit to the
high--ionization coronal lines in the Circinus galaxy using a single
model (i.e., a single continuum shape). From Fig. 5, the best--fit
high-ionization coronal line ratios for the Circinus galaxy correspond to
$-0.8 \leq$ Log~U $\leq 0.8$ for the attenuated, non-thermal continuum
(Model NTA), with an average value Log~U $= 0$. From Fig. 3(a) we
estimate the electron density to be $10^{2} \leq$ Log N$_{e} \leq 10^{4}$
cm$^{-3}$, with an average value, N$_{e} = 10^{3}$ cm$^{-3}$. This
electron density is similar to that derived by Moorwood et al. (1996)
from the density sensitive [\ion{Ne}{v}]24.3($\mu$m)/14.3($\mu$m) ratio,
which gives N$_{e} = 5000$ cm$^{-3}$. We note that the ionization
parameter, Log U $= 0$ quoted above and obtained from the best fit NTA
model, is in units of Log~U for Model NT. From Table~1 (Section 3), the
``true'' ionization parameter for Model NTA equals Log~U $= -1.34$.

The ratios of the observed to the modelled (Model NTA) line fluxes as a
function of ionization energy for the different sulphur, silicon, iron
and magnesium lines seen in the Circinus galaxy are presented in Fig. 8.
The systematic deviation from unity (e.g. sulphur) might be due to
peculiar abundances relative to solar. The different iron species from
[\ion{Fe}{vii}] to [\ion{Fe}{xiv}] fit reasonably well using the
collision strengths adopted in CLOUDY. The details of the atomic data
used are discussed in the CLOUDY manual (HAZY, Ferland 1996), the CLOUDY
source code and Ferland et al. (1998)

\section{Discussion}

The observed coronal line ratios indicate a somewhat lower excitation of
the coronal line region in NGC~1068 than in Circinus, except for the
[\ion{Fe}{xi}]/[\ion{Fe}{x}] ratio which is higher in NGC~1068. The
[\ion{Fe}{vii}]/[\ion{Fe}{x}] ratio is 13 times higher and the
[\ion{Si}{ix}]/[\ion{Si}{vi}] ratio 3.3 times lower in NGC~1068 than in
Circinus. Based on these ratios, Marconi et al. (1996) suggested that the
ionization parameter in NGC~1068 is lower and the ionizing continuum
steeper than in Circinus. From our models, the different observed coronal
line ratios in these two galaxies might be caused by different amounts of
ionizing continuum attenuation. We derive similar ionization parameters
in the coronal line region, but the derived electron densities differ by
a factor of 10, being higher in NGC~1068. Thus, the number of ionizing
photons in the coronal line region of the NGC~1068 is ten times higher
than in the Circinus galaxy.

There are several possible sources of attenuation which could influence
the observed coronal line emission. Binette et al. (1997) and others have
considered the effect of attenuation produced within the coronal line
clouds themselves (ionization stratification) and whether matter-bounded
clouds could overlap along a line-of-sight to the nucleus producing
continuum attenuation for ionization-bounded clouds at larger radii
(shadowing within the coronal line region). While some ionization
stratification seems likely, our models suggest that shadowing by gas in
the coronal line region seems unlikely to be a major cause of attenuation
in NGC~1068. Based on the observed luminosity of [\ion{Si}{ix}], $1.1
\times 10^{40}$ erg s$^{-1}$ (Marconi et al. 1996), and a nuclear
continuum illumination-cone opening angle of $\approx 80\degr$ (Unger et
al. 1992), the required coronal line gas covering factor is only $\approx
7 \times 10^{-2}$. For the Circinus galaxy, in contrast, assuming a cone
opening angle also of $\approx 80\degr$, the implied covering factor is
$\approx 1$, consistent with that proposed by Oliva et al. (1994).

We propose that the different, relative attenuation of the continuum
emission illuminating the coronal line regions in NGC~1068 and the
Circinus galaxy might be due to a different geometrical distribution of
the attenuating gas. This effect might be due to the nuclear radio-jet,
which has a different intensity in these galaxies. Capetti et al. (1997)
found a morphological connection between the optical NLR emission and the
radio emission in the region near to the jet in NGC~1068. The NLR
material along the radio jet is denser and has a higher ionization-state
than the surrounding gas. Axon et al. (1998) also suggest that continuum
anisotropy could be produced by azimuthal variations of the optical depth
to ionizing photons. 

To explain the required mixture of coronal line emission from gas
illuminated by both attenuated and unattenuated continua in NGC~1068, we
propose that the radio jet has swept out a channel through the coronal
line region out into the NLR, pushing some gas aside as it went. Gas
around the channel sees attenuated, non-thermal (nuclear) continuum,
whereas coronal line gas within the channel sees unattenuated,
non-thermal continuum. Thus, in this model, the coronal line emission in
NGC~1068 comes from regions both surrounding the channel around the jet
axis and from within the channel itself. Most of the coronal line
emission comes from close to the nucleus, within $\approx 1\arcsec$.

The Circinus galaxy has no very prominent radio jet near to the nucleus
similar to that in NGC~1068, but it has peculiar, highly--polarized (up
to 45\%) radio lobes orthogonal to the galactic plane, which show a
non-thermal spectrum, and features which may be evidence for outflow
(Elmouttie et al. 1995). An [\ion{O}{iii}] ionization cone was also found
by Marconi et al. (1994). However, the \ion{Fe}{ii} 1.64($\mu$m) and
H$_{2}$ 2.12($\mu$m) images show only weak extended H$_{2}$ emission in
Circinus, suggesting that the jet is confined close to the nucleus
(Davies et al. 1998). The observed blue-shift of the coronal lines
($\approx 35$ km s$^{-1}$; Oliva et al. 1984) in Circinus is also much
smaller than in NGC~1068 ($\approx 300$ km s$^{-1}$; Marconi et al.
1996), consistent with the radio jet having a smaller effect in Circinus.
Thus, it may be that the radio jet has been unable to clear a channel in
the Circinus galaxy, and the coronal line emission comes only from gas
illuminated by the attenuated, non-thermal (nuclear) continuum. 

\section{Conclusions}

We have presented photoionization models for the coronal line spectrum
in NGC~1068 and the Circinus galaxy. The line fluxes were calculated
using four different ionizing continua: a non-thermal (nuclear) continuum
(Model NT); the non-thermal continuum plus a UV bump (Model NTS); and
attenuated versions of both these continua (Models NTA and NTSA). The
attenuation of the continua was assumed to be caused by gas with column
density $10^{22}$ cm$^{-2}$ located within the inner-radius of the
coronal line region. 

For NGC~1068 the predicted [\ion{S}{ix}]/[\ion{S}{viii}],
[\ion{Si}{ix}]/[\ion{Si}{vi}] and [\ion{Fe}{vii}]/[\ion{Fe}{x}] ratios
agree best with the observations if about 40\% of the line emission comes
from gas illuminated by unattenuated, non-thermal continuum (Model NT)
and about 60\% from gas illuminated by attenuated, non-thermal continuum
(Model NTA). The derived electron density and ionization parameter of the
coronal line gas in NGC~1068 are N$_{e} \approx 10^{4}$ cm$^{-3}$ and Log
U $= -1.4$.

For the Circinus galaxy the predicted coronal line ratios fit best if all
the gas is illuminated by attenuated, non-thermal continuum (Model NTA).
The derived ionization parameter, Log~U $= -1.34$, is very similar to
that of NGC~1068. However, the derived electron density of the coronal
line gas in Circinus, N$_{e} = 10^{3}$ cm$^{-3}$, is an order of
magnitude smaller, implying a much higher photon flux compared to
NGC~1068, explaining the different observed excitation state of the
coronal line gas in the two galaxies.

We propose that the difference between the coronal line emission in these
galaxies could be due to the different prominence of their respective
radio jets. In this model, the stronger radio jet in NGC~1068 has swept
out a channel as it passed through the coronal line region into the NLR,
and therefore the radiation from the non-thermal continuum source suffers
less attenuation along the jet axis. Coronal line gas within the channel
sees unattenuated, non-thermal continuum, whereas coronal line
gas in the region around the channel sees attenuated, non-thermal
continuum. We observed the combined coronal line emission from both these
regions. The less powerful radio jet in the Circinus galaxy has been
unable to sweep out such a channel, and hence its coronal line region 
is illuminated only by attenuated, non-thermal continuum.

\begin{acknowledgements}
L.S.Nazarova wishes to thank the University of Leicester for
their hospitality. We thank Gary Ferland for providing a copy of CLOUDY.
The calculations were performed on SUN and 
DEC workstations provided by the PPARC Starlink project at Leicester.
L.S.N would like to acknowledge support under The Royal Society 
grant. This work was also partly supported 
from INTAS 96-0328 and RBRF 96-02-17625 grants. 
\end{acknowledgements}

\end{document}